# Electronic structure of superconducting nickelates probed by resonant photoemission spectroscopy

Zhuoyu Chen[1,2,3]*, Motoki Osada[1,2,3], Danfeng Li[1,3,4], Emily M. Been[1,5], Su-Di Chen[1,2,3], Makoto Hashimoto[6], Donghui Lu[6], Sung-Kwan Mo[7], Kyuho Lee[1,3,5], Bai Yang Wang[1,3,5], Fanny Rodolakis[8], Jessica L. McChesney[8], Chunjing Jia[1], Brian Moritz[1], Thomas P. Devereaux[1,3,9], Harold Y. Hwang[1,2,3], Zhi-Xun Shen[1,2,3,5]*

[1]Stanford Institute for Materials and Energy Sciences, SLAC National Accelerator Laboratory, Menlo Park, CA 94025, USA.

[2]Department of Applied Physics, Stanford University, Stanford, CA 94305, USA.

[3]Geballe Laboratory for Advanced Materials, Stanford University, Stanford, CA 94305, USA.

[4]Department of Physics, City University of Hong Kong, Kowloon, Hong Kong

[5]Department of Physics, Stanford University, Stanford, CA 94305, USA.

[6]Stanford Synchrotron Radiation Lightsource, SLAC National Accelerator Laboratory, Menlo Park, California 94025, USA.

[7]Advanced Light Source, Lawrence Berkeley National Laboratory, Berkeley, California 94720, USA

[8]Advanced Photon Source, Argonne National Laboratory, Lemont, Illinois 60439, USA.

[9]Department of Materials Science and Engineering, Stanford University, Stanford, California 94305, USA.

*Correspondence to: zychen@stanford.edu, zxshen@stanford.edu

Lead contact: Zhi-Xun Shen (zxshen@stanford.edu)

**Summary**

The discovery of infinite-layer nickelate superconductors has spurred enormous interest. While the $Ni^{1+}$ cations possess nominally the same $3d^9$ configuration as $Cu^{2+}$ in cuprates, the electronic structure variances remain elusive. Here, we present a soft x-ray photoemission spectroscopy study on parent and doped infinite-layer Pr-nickelate thin films with a doped perovskite reference. By identifying the Ni character with resonant photoemission and comparison to density functional theory + U (on-site Coulomb repulsion energy) calculations, we estimate U ~ 5 eV, smaller than the charge transfer energy Δ ~ 8 eV, confirming the Mott-Hubbard electronic structure in contrast to charge-transfer cuprates. Near the Fermi level ($E_F$), we observe a signature of occupied rare-earth states in the parent compound, which is consistent with a self-doping picture. Our results demonstrate a correlation between the superconducting transition temperature and the oxygen 2p hybridization near $E_F$ when comparing hole-doped nickelates and cuprates.

**Introduction**

After decades of quest, the nickel analogs of the cuprate superconductors have been recently found, shedding new light on the path towards understanding the origin of high-temperature superconductivity.[1–3] Although originally the superconducting nickelates with infinite layer structure were designed through mimicking the $3d^9$ configuration of the undoped cuprates,[4] recent theoretical[5–17] and experimental[18–20] studies of the nickelate electronic structures imply important differences from the cuprates. Direct experimental probes of the electronic structure of the nickelates have become the central need.[21] Photoemission spectroscopy provides a measurement of the density of states (DOS) over a wide range of binding energies by analyzing photoelectrons emitted from the material surface.[22] As a highly surface sensitive tool, its application to the nickelate superconductors is challenging at the current stage. The infinite-layer phase of superconducting nickelates is formed through epitaxial growth of the perovskite phase, followed with a soft chemistry topotactic reduction. A top capping layer of $SrTiO_3$ is usually used for the coherent quality of the nickelate films, particularly for the top layers[23]. Therefore, the difficulty of photoemission measurements is two-fold: first, the necessary *ex situ* reduction process introduces surface exposure to the atmosphere; second, the few-unit-cell-thick capping layer impedes the sub-nanometer level probe depth of usual angle-resolved photoemission experiments.

To mitigate these obstacles, we conduct photoemission spectroscopy (PES) measurements, utilizing the soft x-ray range of photon energy, on the newly developed Pr infinite-layer nickelate thin films (i.e. the parent compound $PrNiO_2$ and superconducting doped compound $Pr_{0.8}Sr_{0.2}NiO_2$ with critical temperature $T_C \sim 12$ Kelvin) that preserve a stable surface during the reduction without capping,[24,25] together with a perovskite $Pr_{0.8}Sr_{0.2}NiO_3$ thin film for reference. The relatively longer probe depth from the higher photoelectron kinetic energy emitted by soft x-ray helps alleviate surface sensitivity. The surface free from capping enables a direct measurement of the DOS of the nickelate films without the influence of the photoemission signal from the cap.

**Results**

Figure 1A shows the measurement configuration of our experiment. Incident photons with circular polarization is angled 35º to grazing, such that both in-plane and out-of-plane electron orbitals can be excited. The photoelectron analyzer is positioned near the normal direction of the thin film surface. The current to the ground is measured simultaneously for the total electron yield of x-ray absorption spectroscopy (XAS). Figure 1B, and C exhibits the XAS for Ni $L_3$ and O K edges, respectively. On the Ni $L_3$ edge, the $Pr_{0.8}Sr_{0.2}NiO_3$ sample exhibits a relative peak shift and a different line shape compared to the infinite layer samples. This is expected for a different Ni oxidization state and consistent with previous results on La nickelates.[18] The $Pr_{0.8}Sr_{0.2}NiO_2$ and $PrNiO_2$ samples shows largely similar line shape. The secondary peak for the doped sample appears slightly higher than the parent sample, which is a manifestation of doped holes in the Ni $3d$ orbitals.[19,26] On the O K edge, the suppression of the strong pre-peak in the $Pr_{0.8}Sr_{0.2}NiO_3$ sample after reduction reflects the shift of O partial DOS away from the Fermi level for both the infinite-layer samples[18]. Note that the XAS data are collected based on the total electron yield, which is generally more sensitive to near-surface region compared to fluorescence yield mode[18]. Figure 1D shows the x-ray photoelectron spectroscopy (XPS) for O K. These O K peaks have similar peak width with a relative shift between one another. Relative to $PrNiO_2$, the peaks for $Pr_{0.8}Sr_{0.2}NiO_2$ and $Pr_{0.8}Sr_{0.2}NiO_3$ shift around 0.2 eV and 0.6 eV, respectively. These peak shifts are likely due to chemical potential shifts. Additional XPS and XAS data are presented in Figs. S1 and S2.

To probe the chemical potential shifts between these three samples, PES spectra of 260 eV photon energy for the valence band DOS are shown in Fig. 2A. Photons of 1360 eV, in principle corresponding to significantly larger probing depth (~2 nm ≈ 6 unit cells) compared to 260 eV (~0.7 nm ≈ 2 unit cells), lead to basically identical PES features with slightly larger broadening (displayed in Fig. S3), indicating that the 260 eV spectra are representative of intrinsic bulk properties. While the PES spectral line shape of the perovskite phase sample $Pr_{0.8}Sr_{0.2}NiO_3$ is quite different, the two infinite-layer-phase spectral curves are almost identical with 0.2 eV of overall energy shift. If we use the midpoint (intensity = 0.5, circles in Fig. 2A) as a representative energy of the valence band top, $Pr_{0.8}Sr_{0.2}NiO_3$ has a 0.6 eV energy shift with respect to $PrNiO_2$. The same energy shift is also observed for the peaks of the valence band DOS (diamonds in Fig. 2A). The 0.2 eV and 0.6 eV values are consistent with the O K core level XPS peak shifts (Fig. 1D); thus, we attribute them to chemical potential changes coming from hole doping.

To further analyze the DOS, we compare the $PrNiO_2$ experimental data with density functional theory calculation for different on-site Coulomb repulsion energy U values (DFT+U),[9] as shown in Fig. 2B. By varying U, the Ni 3$d$ partial DOS envelop (blue) continually shifts to higher binding energy, while the O 2$p$ partial DOS envelop (red) remains mostly at the same position with some spreading towards $E_F$ near the high-U end. When U = 0 (non-interacting), the upper tail of the Ni 3$d$ envelop crosses $E_F$, but a gap is formed when U = 3 eV. The overlapping between Ni and O is small when U is low and becomes significant when U is larger than 6 eV. By Gaussian peak fitting, we found there are two prominent features in the experimental DOS, matching the Ni 3*d* and O 2*p* partial DOS envelops found in the DFT+U simulations. The Ni partial DOS correspondence is justified by resonant photoemission data, to be discussed in Fig. 3. The binding energies of prominent features of the simulated partial DOS are quantified and plotted as a function of U in Fig. 2C. The comparison of the Ni feature with experimental fitted center of weight values gives an estimate of U = (5 ± 1) eV. The error is estimated based on an analysis of the overlap between experimental and simulated Ni partial DOS as a function of U, as discussed in the Fig. S4. This extracted value of U is consistent with recent resonant inelastic x-ray scattering studies[20] and theorectical predictions based on seven-orbital models.[13,27]

With the understanding of the infinite-layer electronic structure, here we discuss the differences among cuprates, infinite-layer nickelates, and perovskite nickelates, as illustrated in Fig. 2D. While cuprates[4] are known to be charge transfer insulators where the charge transfer energy $\Delta$ is far smaller than the Hubbard U, infinite-layer nickelates are in the opposite regime, having U ~ 5 eV and $\Delta \sim U + (E_{Ni3d} - E_{O2p}) \approx 8$ eV, where $E_{Ni3d}$ and $E_{O2p}$ are the experimental fitted binding energies of the Ni 3*d* and O 2*p* features (Fig. 2C), respectively. For perovskite nickelates, previous studies[28–31] have shown that it is effectively a "negative charge transfer" compound with dominating O 2*p* states crossing $E_F$: the significant number of empty O 2*p* states above $E_F$ gives rise to the strong pre-peak in XAS near the O K edge (Fig. 1C). As such, contrasting to cuprates and to a lesser extent perovskite nickelates, the states near the Fermi level in infinite-layer nickelates have dominating 3*d* character.

To clarify the elemental specificity of states observed in the photoemission spectra, we performed resonant PES measurements near the Ni $L_3$ absorption edge, which enhances spectral intensity from Ni 3*d* states through additional emission channels allowed in the resonance process. Figure 3A shows an example comparison of the detected photoemission signals between the off- and on-resonant cases. The off-resonant (before the XAS edge, black, corresponding to #0 photon energy shown in Fig. 3C) curve represents a typical photoemission spectrum, with both valence band (consistent with Fig. 2A) and Pr 5*p* states observed. The on-resonant curve (post-edge, green, corresponding to #4 in Fig. 3C) apparently has enhanced intensity. The increased part of the intensity, as shown by the thick purple curve (#4 – #0), comes purely from the additional channels related to Ni absorption, and thus represent Ni response only. This is verified by a complete absence of the Pr 5*p* feature in the subtracted curve.

However, as illustrated in Fig. 3B, this pure Ni response also has two different origins. After absorption, an electron from the Ni 2*p* core level is excited to an empty state above $E_\text{F}$, leaving a hole in the core level. Following this step, there are two possibilities: 1) the excited electron refills the core hole and transfers the energy to a valence electron, which is then emitted with a final state being the same as for a regular photoemission process – the interference of these two channels gives rise to a resonance effect; 2) an electron in an occupied state fills the core hole and energy is transferred to another electron in an occupied state – the emitted electron has a kinetic energy independent of photon energy, thus is an Auger electron rather than a photoelectron.[32] We need to separate the signals with these two different origins to obtain resonant photoemission spectra. Figure 3C displays the photoemission intensity as a function of binding energy with different photon energies. Auger signals have constant kinetic energy, and therefore form diagonal features in Fig. 3C, distinct from photoemission signals with constant binding energy. Since resonant PES and Auger channels start simultaneously following the absorption, signals from both channels overlap on top of each other for the spectra with photon energies before and on the peak of the absorption, making it difficult to resolve individual contributions. For better separation of these two parts, we choose to analyze spectra with photon energy higher than the absorption peak (#1 to #6), such that Auger signals move deeper in the binding energy axis and separate away from useful resonant PES signal near $E_F$. Gaussian peak fitting analysis is shown in Fig. 3d, where one can see clear partition of features with constant and varying binding energies, representing Ni 3$d$ states (blue) and Ni $L_3M_{4,5}M_{4,5}$ Auger electrons (green), respectively.[33] The identification of Ni 3$d$ states here verified the assignment of orbital characters in Fig. 2. Other results of resonant PES are shown in Fig. S5. In particular, resonant PES enhancements observed at Pr $M_{4,5}$ edges indicate the finite hybridization of Pr 4$f$ orbitals in the valence bands. Yet, the recent finding of superconducting $(La,Sr)NiO_2$ confirms the irrelevance of 4$f$ states to superconductivity.[34]

Now we turn to the near-$E_F$ features. Figure 4A shows the higher-statistics photoemission spectra close to $E_F$. Comparing with the perovskite compound, a significant suppression of spectral intensity near $E_F$ is observed in the infinite-layer samples. As discussed earlier, near-$E_F$ states for the perovskite compound are dominated by O 2$p$ orbitals. A prominent "step" feature (between -0.5 eV and $E_F$) observed here is consistent with the observation of a strong pre-peak in the XAS near O K edge in Fig. 1C, both of which come from O 2$p$, but the former represents occupied states while the latter corresponds to unoccupied states. These observations support the effective "negative charge transfer" picture for the perovskite nickelate. In the infinite-layer parent compound, analysis above have shown Ni 3$d$ states are pushed away from $E_F$ due to finite U. An overall analysis of the DFT+U simulation of DOS near $E_F$, with the Pr states and cross sections[35] considered, is shown in Fig. 4B. As U increases from zero, the Ni 3$d$ intensity drops significantly. On the other hand, weakly interacting three-dimensional Pr 5$d$ states have small but finite intensity crossing $E_F$, and dominates the intensity near U ~5 eV. Experimentally, a very small but finite "Fermi step" is indeed observed near $E_F$ for the parent compound (inset of Fig.

4A). With the same measurement geometry, this small feature is depleted by Sr hole-type doping. This indicates that it is an electron-type pocket and must be related to the Pr $5d$ orbital, as illustrated in Fig. 4C. The existence of Pr $5d$ states near $E_F$ presents a sharp contrast to the cuprates, in which rare-earth states are far above $E_F$. The ultra-low spectral intensity for the doped infinite-layer sample might have a few origins: 1) Ni $3d$ has low spectral weight near $E_F$ with even 0.2 eV shift of chemical potential as pointed out by DFT+U simulation; 2) the cross section of Ni $3d$ is significantly lower than rare-earth $5d$.

**Discussion**

Previous reports[2,3,25,34,36] point out that both parent and Sr-doped infinite-layer nickelates exhibit metallic behavior with similar order of magnitude conductivity, in contrast to cuprates where 20%-hole doping gives rise to ~ 100 times enhancement of conductivity.[37] While the Sr doping provides hole-type carriers to the system, the parent infinite-layer nickelate is "self-doped" because of the small electron pocket[7,38] (unlike the parent perovskite nickelate also being "self-doped" via negative charge transfer[28]). Upon doping, the rare-earth states are depleted and simultaneously holes in Ni $3d$ orbital lead to $3d^8$ configuration.[26,38] Intriguingly, the "self-doped" parent infinite-layer compound with the existence of additional rare-earth $5d$ channel manifests weaker superconducting instability compared to the Sr-doped compound,[34] despite the apparent higher DOS (inset of Figure 4A). This possibly suggests the proximity to weakly-interacting rare-earth $5d$ states may reduce the pairing strength.

Recent x-ray scattering measurement exhibits that the nickelates has a comparatively strong magnetic superexchange,[20] in contrast to its significantly lower hole-doped $T_C$, compared with cuprates.[39] This implies the $T_C$ in these unconventional superconductors may have contributions from different sources, in addition to the superexchange. Considering the rapidly developing optimization of the nickelate thin film materials, disorder may play a role for $T_C$ at this stage.[34] Another possible origin may be related to the weaker hybridization with the O $2p$ orbitals in the nickelates near $E_F$ that could reduce the coupling to O-state-associated bosonic degrees of freedom, such as phonons. Interestingly, recent photoemission experiments in one-dimensional cuprate chains found an anomalously strong near-neighbor attraction, beyond what could be induced by superexchange.[40] This attraction, which likely originates from long-range electron-phonon coupling,[41] is compatible with neighboring in-plane electron pairs.

In conclusion, although the infinite-layer nickelates and cuprates are isostructural and have nominally the same $3d^9$ configuration, we have shown evidence that their electronic structures are dissimilar in two aspects. First, the nickelates have smaller U than Δ, making the $NiO_2$ planes the Mott-Hubbard type with states near $E_F$ of mainly Ni $3d$ character, while cuprates as charge transfer insulators are the opposite, possibly suggesting a correlation between the coupling to O $2p$ associated bosonic degrees of freedom and higher $T_C$. Second, in addition to the strongly correlated $NiO_2$ planes, weakly-interacting rare-earth $5d$ states near $E_F$ induce self-doping in the parent compound and play an essential role for conductivity but a complex role for superconductivity.

## Experimental Procedures

### Resource availability

*Lead Contact*

Zhi-Xun Shen, Email: zxshen@stanford.edu

*Materials Availability*

**Sample preparation**

Infinite-layer nickelate films of 6 nm were synthesized by pulsed laser deposition, followed by a subsequent topotactic reduction reaction. It has been found that an upper $SrTiO_3$ cap was unnecessary for stabilizing a uniform single-crystalline infinite-layer structure for Pr-based nickelates, possibly due to a better lattice matching to the $SrTiO_3$ substrate and a larger tolerance factor[24,25]. Therefore, precursor perovskite films were synthesized without an $SrTiO_3$ cap at 570 °C substrate temperature. The laser fluence and oxygen pressure for the growth of $PrNiO_3$ and $Pr_{0.8}Sr_{0.2}NiO_3$ are 1.39 J/cm², 200 mTorr and 2.19 J/cm², 250 mTorr, respectively. The laser repetition was 4 Hz. The perovskite film measured here is 9 nm thick. The topotactic reduction is done by sealing $CaH_2$ and the precursor perovskite film in a glass tube and heating in a tube furnace to 240 °C for 60 min. More details about synthesis and characterizations are reported earlier[24,25].

**Spectroscopy measurements**

Photoemission spectroscopy, XPS, and XAS measurements were performed at the Advanced Photon Source (APS) Sector 29-ID angle-resolved photoemission spectroscopy end station. Samples prepared at Stanford were sealed within vacuum glass tubes for shipment to APS. At the beamline, samples were mounted using tantalum foils in air and loaded into the measurement chamber. We found pre-annealing in ultra-high vacuum (UHV) below 250 °C (temperatures higher than this could lead to sample decomposition) before measurement does not give noticeable improvement to spectroscopic data quality. The measurement temperature for the data shown is lower than 20 Kelvin. We did not observe obvious angle-dependent dispersion of photoemission intensity for the infinite-layer samples, probably due to momentum-disruptive scatterings associated with surface adsorbates from *ex situ* sample handling/treatment processes.

**Theoretical calculations**

Antiferromagnetic DFT+U calculations were performed in two-nickel unit cells with spin resolution, using the generalized gradient approximation (GGA) method and the simplified version from Cococcioni and de Gironcoli[42], as implemented in QUANTUM ESPRESSO[43]. Similar methods were used by Been et al.[9]

**Acknowledgements**

We thank Wei-Sheng Lee, Haiyu Lu, Yufeng Liang, and Yao Wang for discussions. The work at SLAC and Stanford was supported by the U.S. Department of Energy (DOE), Office of Basic Energy Sciences, Division of Materials Sciences and Engineering (contract No. DE-AC02-76SF00515). We acknowledge the Gordon and Betty Moore Foundation's Emergent Phenomena in Quantum Systems Initiative (grant No. GBMF9072) for synthesis equipment. This research used resources of the Advanced Photon Source, a U.S. Department of Energy (DOE) Office of Science User Facility at Argonne National Laboratory and is based on research supported by the U.S. DOE Office of Science-Basic Energy Sciences, under Contract No. DE-AC02-06CH11357. The work at LBNL was supported by DOE under contract No. DE-AC02-05CH11231. Part of the theoretical calculations used resources of National Energy Research Scientific Computing

Center (NERSC), operated under Contract No. DE-AC02-05CH11231. Danfeng Li acknowledges the financial support from City University of Hong Kong under Project 9610500.

**Author Contributions**

**Declaration of Interests**

The authors declare no competing interests.

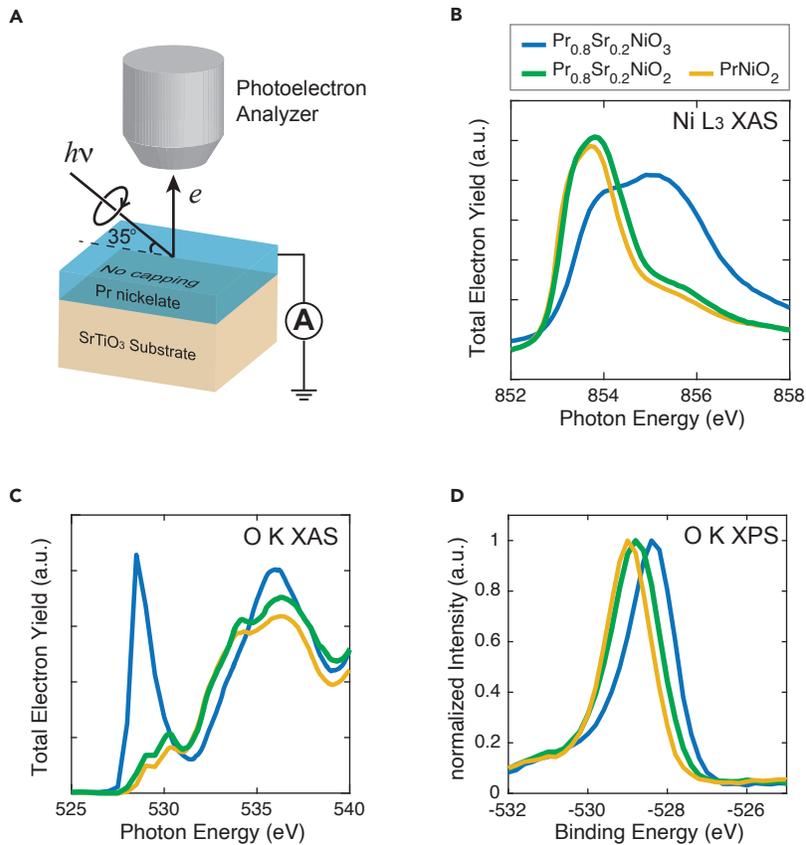

**Figure 1. X-ray absorption (XAS) and x-ray photoelectron spectroscopy (XPS)**

(A) Schematic diagram of the experimental setup.

(B) XAS spectra of the three different samples near the Ni L3 edge. Assuming the rising edges in both infinite layer compound correspond to $2p^63d^9$–$2p^53d^{10}$ transition, the $Pr_{0.8}Sr_{0.2}NiO_2$ curve is slightly scaled and shifted in intensity to match the $PrNiO_2$ rising edge for comparison.

(C) O K edge XAS.

(D) O K XPS peaks of the three samples.

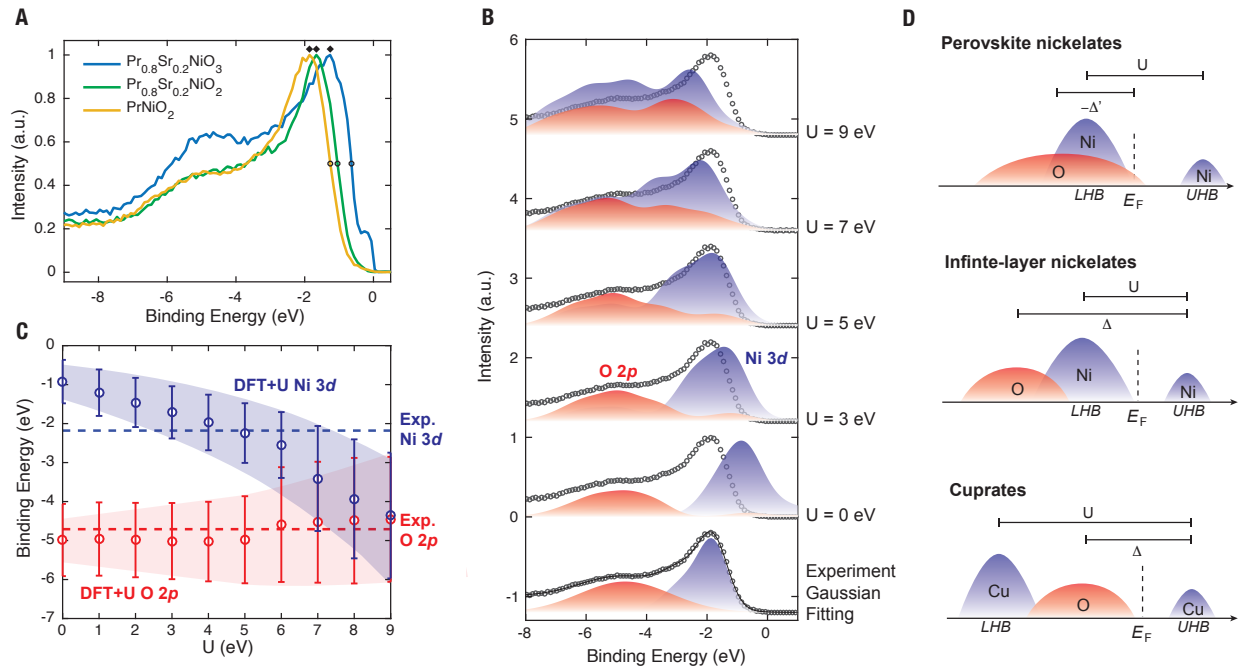

**Figure 2. Density of states**

(A) Photoemission spectra of the three samples with maximum intensity normalized to one. Diamonds and circles represent peak and midpoint positions, respectively. Measurement photon energy is 260 eV.

(B) DFT+U results with different representative U values compared with Gaussian fitting of experiment. Grey circles are experimental data for $PrNiO_2$. Black solid curve is the Gaussian fitting. Blue shaded envelops represent Ni $3d$ related states. Note that, for the experiment, the blue shaded envelop consist of two fitted Gaussian peaks. Red shaded envelops represent O $2p$ related states. The simulated shaded envelops are with 0.5 eV broadening.

(C) Binding energies of the Ni $3d$ and O $2p$ prominent features (center of weight, only features with > 40% of peak intensity are counted) as a function of U, based on DFT+U simulation. Error bars represent standard deviations. Dashed lines correspond to values from Gaussian fitting of experiment.

(D) Schematic diagrams of the electronic structure of cuprates, infinite-layer nickelates, and perovskite nickelates.

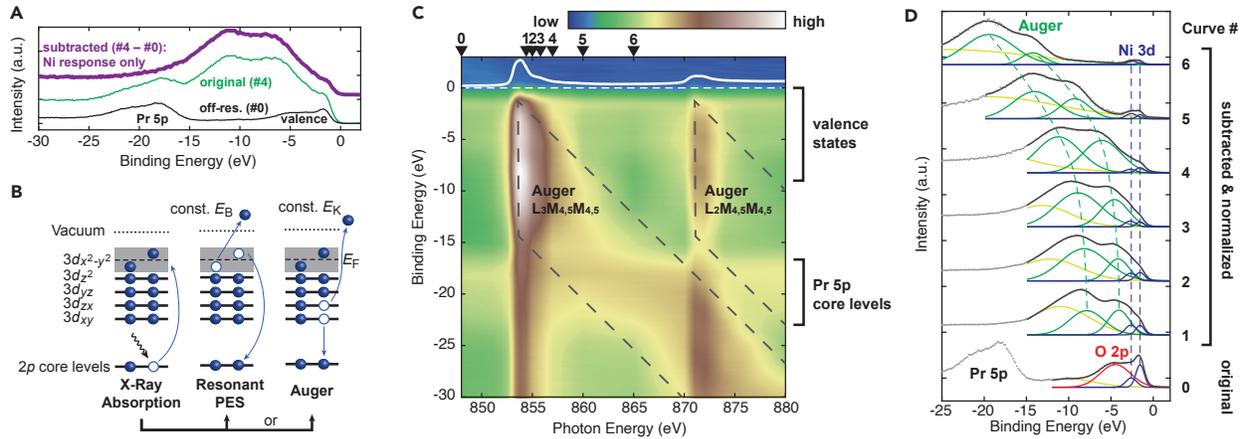

**Figure 3. Resonant photoemission spectra at Ni edges**

(A) Spectra for photon energies of 848 eV (black, off-resonant, corresponding to #0 in (C)), 857 eV (green, absorption enhanced, corresponding to #4 in (C)), and the subtracted spectrum (#4 – #0). The subtracted curve, shifted up for clarity, represent response from Ni states only.

(B) After the $2p^63d^9$–$2p^53d^{10}$ absorption, both resonant photoemission and Auger channels bring the system to $2p^63d^8$. The Auger end configuration shown here is one example out of many possibilities. The kinetic energy of the resonant photoemitted electron increase proportionally with photon energy increase, such that binding energy is constant like regular photoemission process, while the kinetic energy of the Auger electron is invariant of photon energy change.

(C) Spectra at different photon energies in color plot. Areas marked by dashed lines corresponds to Auger dominant regions.

(D) Spectra with Gaussian fitting analysis for photon energies marked by #0 to #6 shown with triangles in (C). Blue, red, and green curves represent Ni states, O states, and Auger features, respectively. Grey dots are experimental data. #1 to #6 curves shown here are after subtracted with #0 curve and normalized with the maximum intensity for clarity of presentation. The #0 off-resonant curve is original, same as that shown in (A).

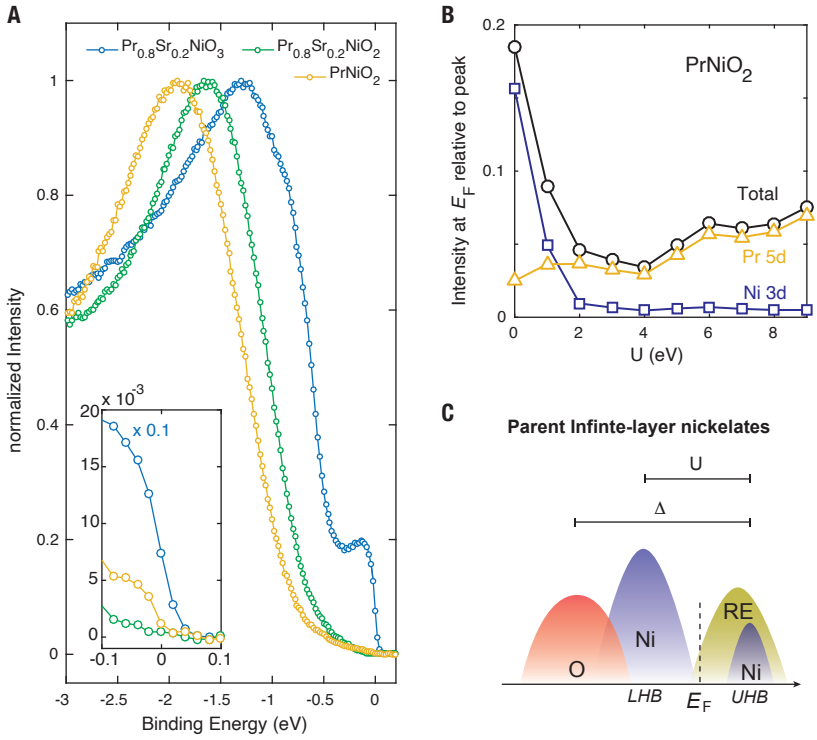

**Figure 4. Near-$E_F$ features**

(A) Photoemission spectra near $E_F$ of the three samples with maximum intensity normalized to one. Inset is a magnification of a small region around $E_F$. Measurement photon energy is 260 eV.
(B) DFT+U simulation results of intensity at $E_F$ of different orbitals as a function of U with cross sections considered. The broadening utilized here is 34 meV.
(C) Schematic diagram of electronic structure for parent infinite-layer nickelates. RE: rare earth.

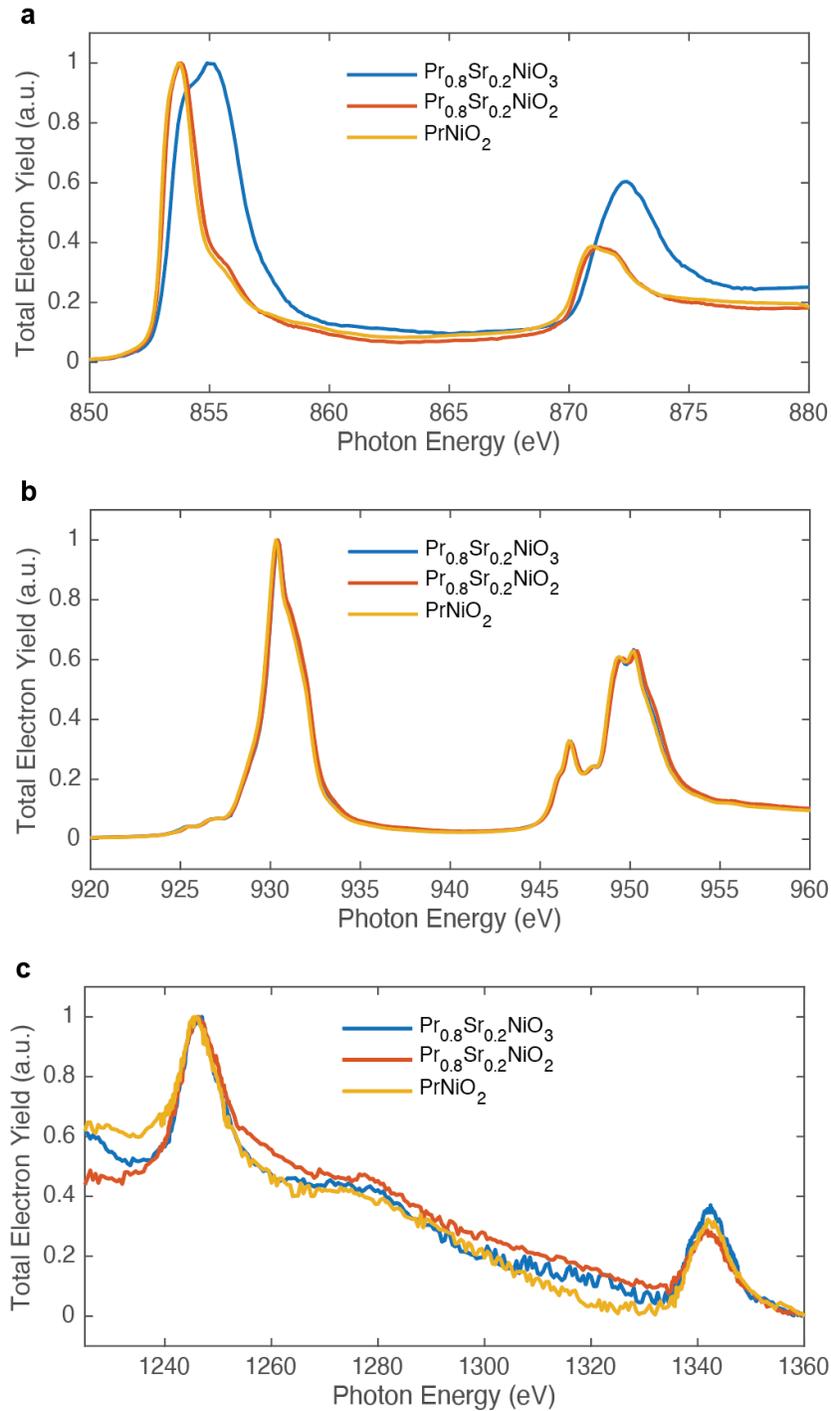

**Figure S1**. **X-ray absorption spectroscopy (XAS).** XAS spectra of the three samples near Ni $L_{2,3}$, Pr $M_{4,5}$, and Pr $M_{2,3}$ edges are shown in panels **a**, **b**, and **c**, respectively. The total electron yields are all normalized for comparison. The Pr $M_{2,3}$ XAS peaks are broad and weak such that the resonant photoemission enhancement for Pr $5d$ states near $E_F$ is below measurement detectable level.

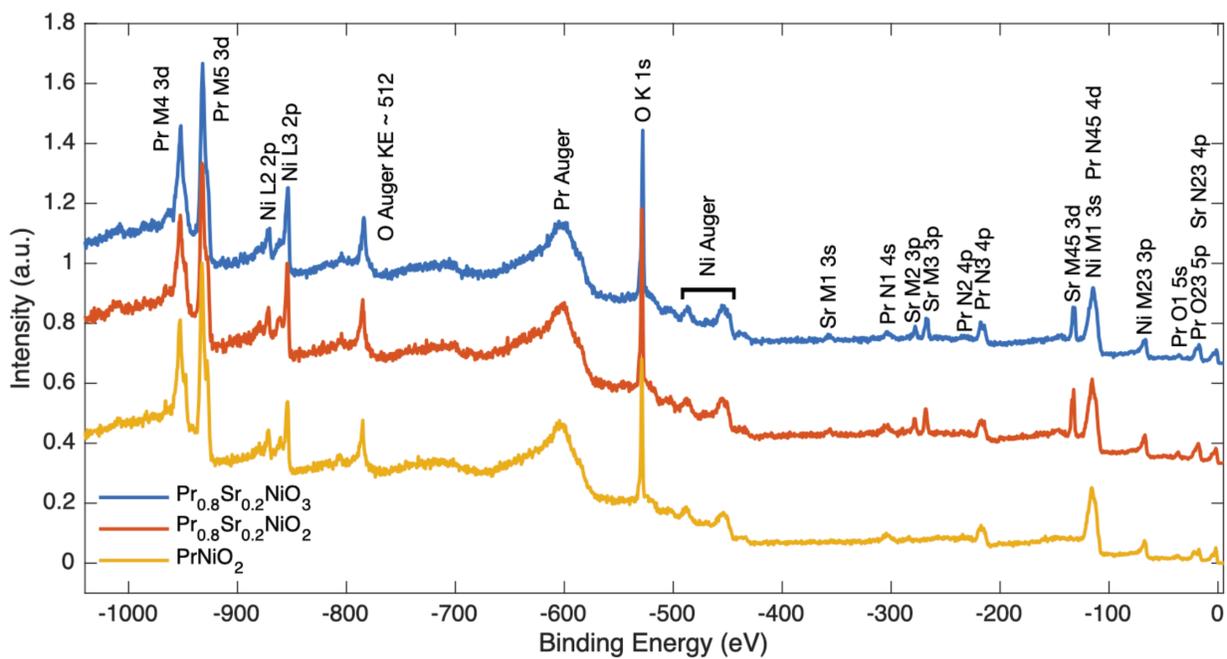

**Figure S2. X-ray photoelectron spectroscopy (XPS).** Wide-range XPS for the three samples measured with photon energy 1300 eV.

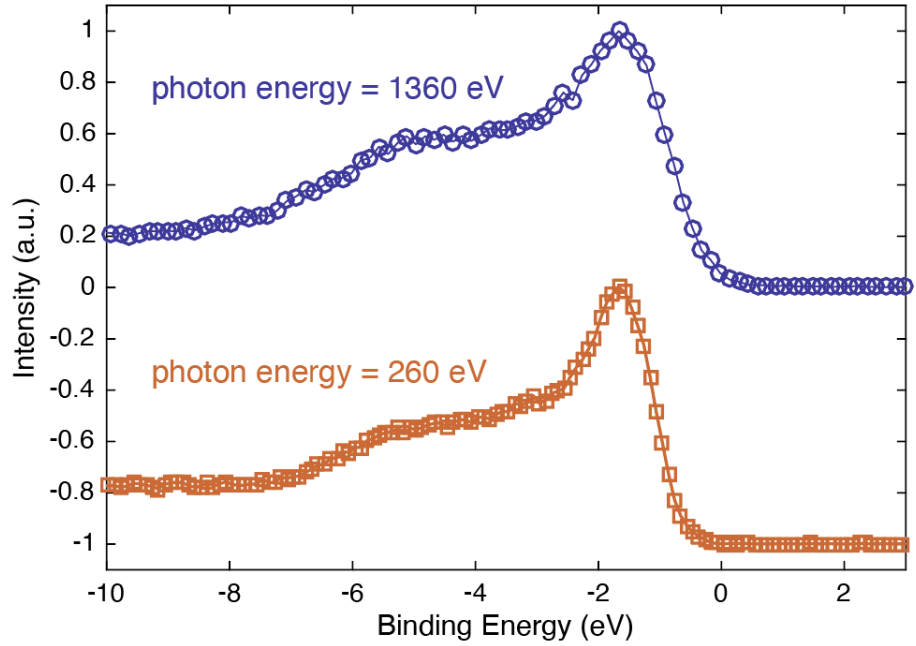

**Figure S3**. **Photon energy independence.** Photoemission spectra of the $Pr_{0.8}Sr_{0.2}NiO_2$ sample with 1360 eV (top) and 260 eV (bottom) photon energies. In principle, the probing depth of the 1360 eV case is around 2 nm, equivalent to about 6 unit cells, which is almost triple of the 260 eV case (~ 0.7 nm ~ 2 unit cells).

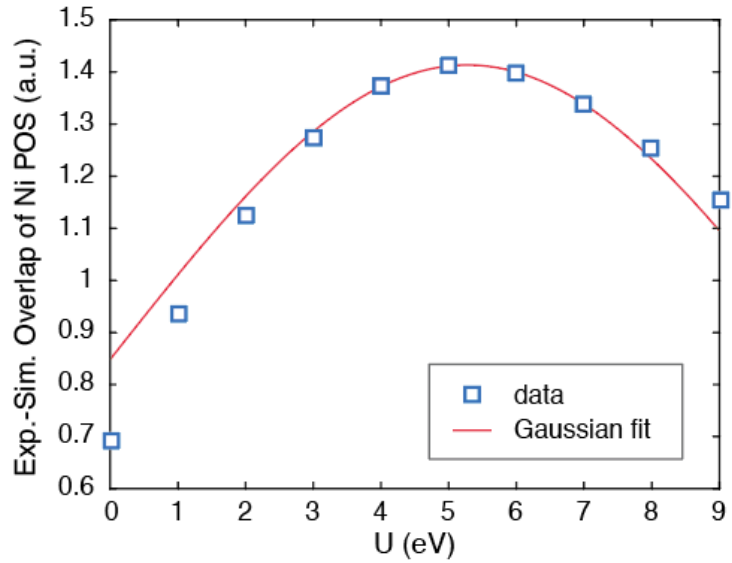

**Figure S4. Error bar estimation.** Overlapped area between experimental Ni partial DOS based on Gaussian peak fitting and DFT+U simulated Ni partial DOS as a function of Hubbard U. Blue squares are data calculated from normalized partial DOS. Red curve is a Gaussian peak fit to the blue squares near the peak. The fitted mean is U = 5.3 eV, with 95% confidence bound being 0.3 eV. We estimate the error to be a combined effect from both the Gaussian fitting and finite data interval: $\Delta U \sim \sqrt{(0.3^2+1^2)} = 1.0$ eV. Thus, we estimate the peak position of this overlap U = (5 ± 1) eV.

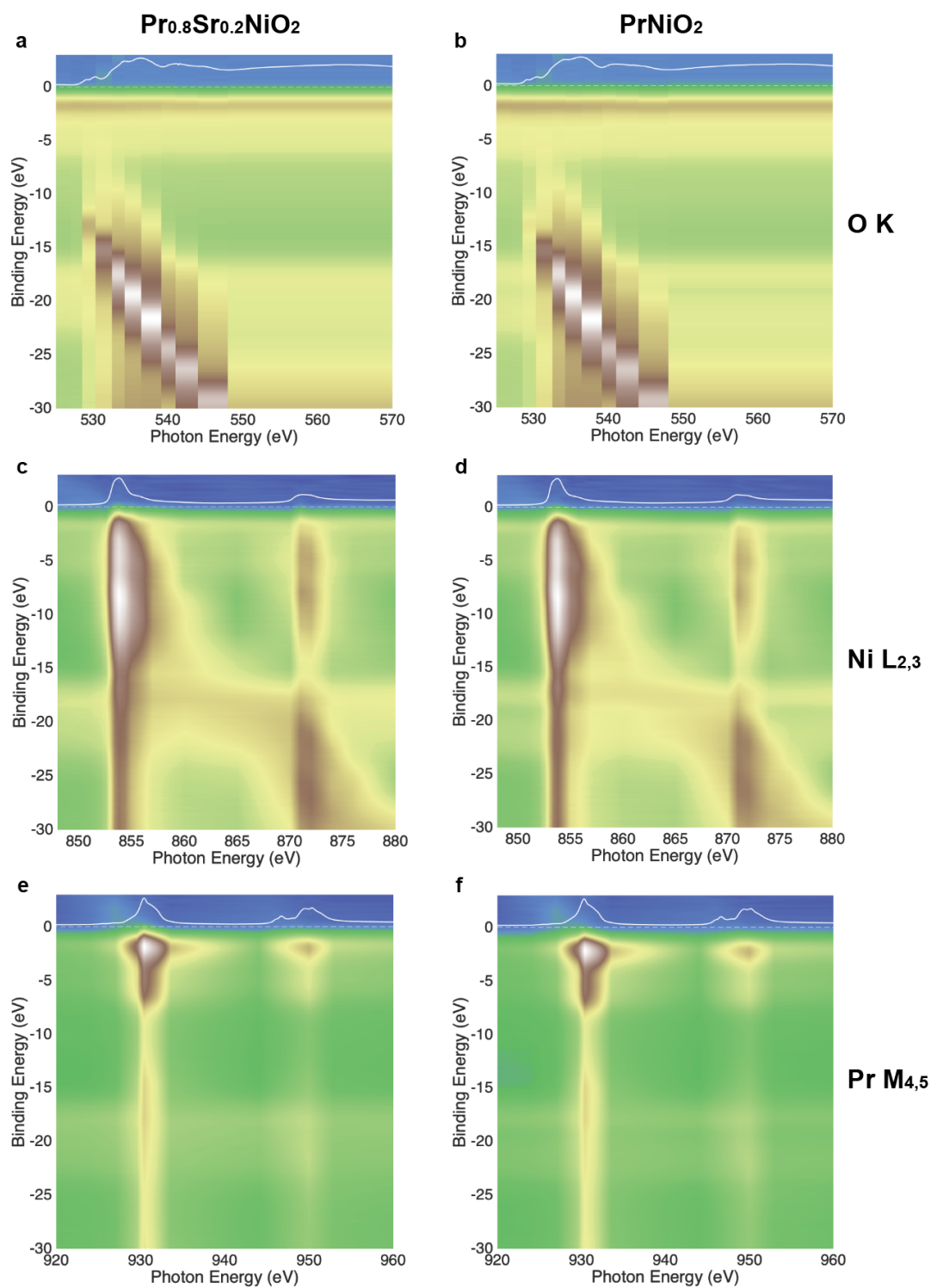

**Figure S5. Resonant PES.** Photoemission spectra of infinite layer nickelates at different photon energies in color plots, for O K (a,b), Ni $L_{2,3}$ (c,d), and Pr $M_{4,5}$ (e,f), respectively. Near O K edge, only Auger signal can be seen. The photoemission channel did not display any detectable resonant effect near O K edge. Near Pr $M_{4,5}$ edge, there is no Auger signal interference. The resonant enhancement near the Pr $M_{4,5}$ edge indicates the existence of Pr $4f$ states in the valence band.

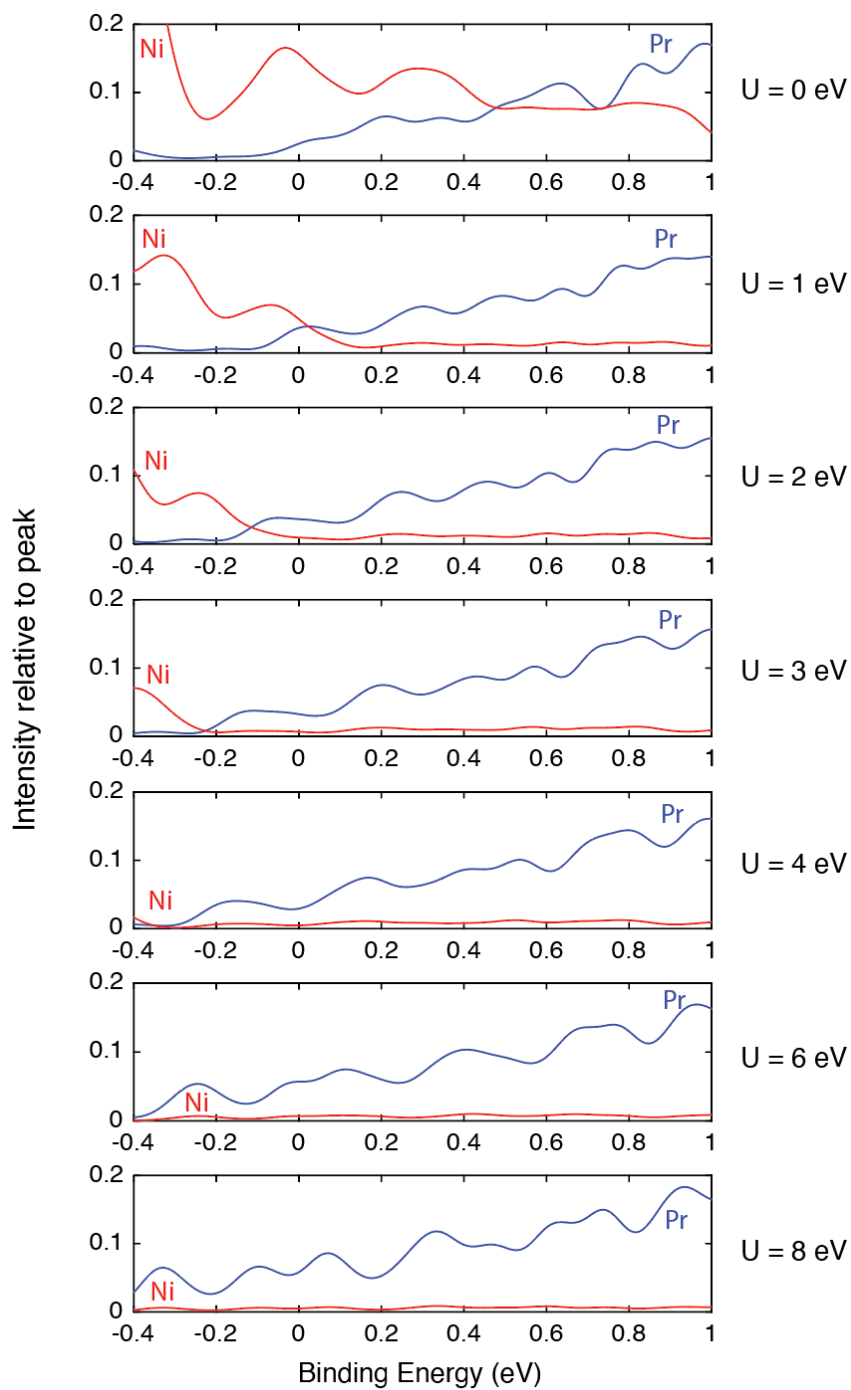

**Figure S6. Numerical calculation results near $E_F$.** DFT+U simulated U-dependent intensity with Pr 5d (blue) and Ni 3d (red) characters relative to the peak total intensity, considering the cross sections.